\documentclass[aps,pre,twocolumn,superscriptaddress]{revtex4}
\usepackage{amsmath,amssymb}
\usepackage{graphicx}
\usepackage{subfigure}
\usepackage{color}

\usepackage{psfig}
\usepackage{amssymb}
\usepackage{amsmath}
\usepackage{times}
\usepackage{tabularx}
\usepackage{color}

\makeindex

\newcommand{\erf}{\ensuremath{\mathop{\rm erf}}}

\newcommand{\vect}[1]{\mathbf{#1}}

\newcommand{\vdiff}[2]{\left|\vect{#1} - \vect{#2}\right|}

\newcommand{\V}{\ensuremath{\mathcal{V}}}

\newcommand{\sig}{\ensuremath{\sigma}}

\newcommand{\vs}{\ensuremath{v_0(r)}}
\newcommand{\vl}{\ensuremath{v_1(r)}}

\newcommand{\rhoqs}{\ensuremath{\rho^{q \sigma}}}
\newcommand{\rhoq}{\ensuremath{\rho^q}}
\newcommand{\rhoG}{\ensuremath{\rho_G}}

\begin{document}

\title{\bf {Acetonitrile on silica surfaces and at its liquid-vapor interface:
structural correlations and collective dynamics}}

\author{Zhonghan Hu}
\altaffiliation[Present address: ]{State Key Laboratory of Supramolecular Structure and Materials, Jilin
University, Changchun 130012, China}
\affiliation{Institute for Physical Science and Technology, University of
  Maryland, College Park, Maryland 20742}

\author{John D. Weeks}
\email{jdw@umd.edu}
\affiliation{Institute for Physical Science and Technology, University of
  Maryland, College Park, Maryland 20742}
\affiliation{Department of Chemistry and Biochemistry, University of Maryland,
  College Park, Maryland 20742}

\begin{abstract}
Solvent structure and dynamics of acetonitrile at its liquid-vapor (LV) interface
and at the acetonitrile-silica (LS)
interface are studied by means of molecular dynamics simulations.
We set up the interfacial system and treat the
long-ranged electrostatics carefully to obtain both stable LV and
LS interfaces within the same system.
Single molecule (singlet) and correlated density orientational profiles and singlet and
collective reorientational dynamics are reported in detail for both
interfaces. At the LS interface acetonitrile forms layers. 
The closest sublayer is dominated by
nitrogen atoms bonding to the hydrogen sites of the silica surface.
The singlet molecular reorientation is strongly hindered when close to
the silica surface, but at the LV interface it relaxes much faster than in the bulk.
We find that antiparallel correlations between
acetonitrile molecules at the LV interface are even
stronger than in the bulk liquid phase. This strong antiparallel correlation
disappears at the LS interface. The collective reorientational
relaxation of the first layer acetonitrile is much faster than the
singlet reorientational relaxation but it is still slower than in the bulk.
These results are interpreted with reference to a variety of
experiments recently carried out. 

In addition, we found that defining interface properties based on
the distribution of positions of different choices of atoms or sites within
the molecule leads to apparently different orientational profiles, especially at the
LV interface.
We provide a general formulation showing that this ambiguity arises when the size of
the molecule is comparable to the interfacial width and is particularly significant when
there is a large difference in density at the upper and lower
boundaries of the interface. We finally analyze the effect of the long
ranged part of the electrostatics to show the necessity of properly treating
long-ranged electrostatics for simulations of interfacial systems.
\end{abstract}
 \maketitle


\section{Introduction}
\label{sec:intro}
Interfacial properties of liquids has been a challenging and important
subject to study because both structure and dynamics of molecules 
at interfaces are usually quite different from those in bulk
liquids. In the past two decades, several experimental groups have focused
on the acetonitrile system confined in silica nanopores or at its
air/liquid interface using a variety of experimental tools
\cite{Nikiel_Zerda1990,Zhang_Eisenthal1993,Zhang_Jonas1993,
Tanaka_Kaneko1998,
Loughnane_Fourkas1998,Loughnane_Fourkas1999,
Loughnane_Fourkas1999a,Kim_Somorjai2003,
Kittaka2005,Kittaka2007}.
Early NMR
experiments by Zhang and Jonas suggested a two-state 
model, which says that the liquids confined in pores with diameter
24 {\rm \AA} or 44 {\rm \AA} consist of both an adsorbed component and a
bulk-like component \cite{Zhang_Jonas1993}.
Additional support for this picture was provided by Fourkas and coworkers,
who employed optical Kerr effect (OKE) spectroscopy to
study the dynamics of liquids confined in silica nanopores. Their
experiments suggested that hydrogen bonding is responsible for the
extreme inhibition of dynamics at the pore
surface \cite{Loughnane_Fourkas1998,Loughnane_Fourkas1999,Loughnane_Fourkas1999a}.
Some shortcomings of the two-state model in describing detailed features
of the dynamics have been noted and
addressed \cite{Loughnane_Fourkas1999a,Morales_Thompson2009}.

Other experimental work using isotherm adsorption
techniques has been explained by dividing liquid acetonitrile molecules
confined in
pores into monolayer acetonitrile and capillary condensed acetonitrile
\cite{Tanaka_Kaneko1998,Kittaka2005}.
Kittaka and coworkers have recently carried out both spectroscopic
and temperature-gradient experiments to determine low temperature
properties of acetonitrile confined in mesoporous material
MCM-41 \cite{Kittaka2005}. Their measurements have shown that the monolayer
acetonitrile adsorbed onto the inner surface of the cylindrical pores
interacts strongly with surface hydroxyls, in agreement with the earlier
finding from the OKE experiment. Their work and the earlier Raman
experimental 
work by Tanaka  and coworkers \cite{Tanaka_Kaneko1998} both pointed out that
the rotational relaxation time scale of the remaining capillary condensed
molecules inside the pores is similar to that of the bulk liquid phase.
They have also found that the melting point in the confined
silica pores from phase transition of $\alpha$-phase crystal to liquid
acetonitrile is much lower than the value measured in the bulk.

Surface sensitive experiments 
have also examined the liquid-vapor (LV) interfacial
properties of acetonitrile and the binary mixture of water and
acetonitrile
\cite{Zhang_Eisenthal1993,Kim_Somorjai2003}. These experiments
aim to gain information on molecular ordering at the interface 
where the molecular environment is highly inhomogeneous.
Using the sum frequency generation technique, Eisenthal and coworkers showed that
the vibrational frequency of acetonitrile changes in mixed solvents as its concentration at
the LV interface is varied \cite{Zhang_Eisenthal1993}.
Their work indicates that acetonitrile
molecules prefer to lie parallel to the interface at higher
concentrations. The gradual change of preferred orientation
was later confirmed by Somorjai and coworkers \cite{Kim_Somorjai2003} but
there is some disagreement between the results of these two
studies \cite{Zhang_Eisenthal1993,Kim_Somorjai2003} concerning
the magnitude of the tilting angle and its dependence on the bulk
concentration. 

A few theoretical and computational studies have 
been carried out to provide a molecular description of the interfacial structure
and dynamics for both the LV and the liquid-silica (LS) interfacial
acetonitrile systems
\cite{Mountain2001,Paul_Chandra2005,Morales_Thompson2009,Gulmen_Thompson2009,Cheng_Berne2009}.
Very recently, Thompson and coworkers carried out molecular dynamics (MD)
simulations of acetonitrile confined in silica pores with varying hydrobocity and analyzed in detail
the structure and dynamics of the confined acetonitrile. Their work has shown 
that confinement and modifications in the hydrogen bond network can lead to line shape broadening
and slowdown of the rotational dynamics \cite{Morales_Thompson2009,Gulmen_Thompson2009}. 
A detailed MD study of the LV interface
using both polarizable and nonpolarizable models of acetonitrile was also
carried out recently by Paul and Chandra \cite{Paul_Chandra2005}. They
concluded that acetonitrile molecules at the interface prefer to orient
with their dipoles parallel to the surface. Their work also indicated that
the relaxation time scale of
molecular rotation at the LV interface is much shorter
than that in bulk solution.

Although both simulations and experiments on
acetonitrile in the silica pores agree that
strong surface-molecule interactions tends to slow down the dynamics of
acetonitrile,
it is not known how much of this effect is due to the cylindrical confinement and 
how properties would change as a function of distance from planar interfaces.
While single molecule (singlet) density, singlet orientational profiles and singlet
rotational dynamics are often analyzed in
simulation studies of interfacial systems, less attention has ben paid to
the structural correlations between pairs of molecules and to
collective dynamics. Thus several questions remain open.
How does the planar silica surface affect liquid correlation as a function of distance from the
surface and how
do the correlations near the silica surface, in the LV
interface, and in the bulk differ from one another?
Are the collective reorientational time scales affected by the
inhomogeneous environment? Do they differ from the corresponding singlet
reorientations?

In order to answer the above questions and to
provide a molecular description of structure, correlation, and singlet and
collective dynamics,
we investigate a planar silica/liquid/vapor acetonitrile system using
classical MD simulations.
The potential of the planar silica surface is taken
from the work by Lee and Rossky \cite{Lee_Rossky1994}. It has also been
used in recent simulations of water confined by silica
planar surfaces by Giovambattista, Rossky and Debenedetti
\cite{Giovambattista_Rossky2006,Giovambattista_Debenedetti2007,Giovambattista_Debenedetti2007a,Giovambattista_Debenedetti2009}.

To better compare our simulation with ongoing experiments of
acetonitrile liquid on a planar silica surface \cite{Fourkas_Walker2009},
we attempt to make the simulation
setup as close as possible to the experimental conditions.
Our simulation differs from
many other MD simulations of interfacial systems in the following ways:
(i) Continuum implicit external potentials are used account for the
long-ranged dispersion interaction from the semi-infinite bulk silica region. (ii)
Long-ranged electrostatics in this slab geometry are treated with the
computationally efficient corrected
Ewald3D method originally developed by Yeh and
Berkowitz \cite{Yeh_Berkowitz1999} to correct for spurious periodic images normal to the 
slab walls. The accuracy of this method is
compared with the rigorous Ewald2D method \cite{Heyes_Clarke1977} for the
dimensions of our system. (iii) Our system setup allows
interfacial, bulk and vapor acetonitrile to coexist in one simulation.
Acetonitrile molecules in the system on average feel nearly zero pressure 
from the boundaries and interfacial molecules can readily exchange with
the bulk region.

The rest of the paper is organized as follows. Details of the surface
construction, a validity check of the potentials, and simulation parameters
are provided in section~\ref{sec:details}. In section~\ref{sec:results} we
investigate density distributions, orientational
profiles, radial angular correlations, and singlet and collective reorientations
in detail to give explanations for a variety of experimental results. The
electrostatic potential and its long-ranged effects are also analyzed in this
section to show their relative importance for both interfaces. We finally
draw our conclusions in section~\ref{sec:con}.

\section{Simulation details}
\label{sec:details}
\subsection{Surface potential and long-ranged electrostatics}
\label{sec:sur} 
We constructed a four-layer silica surface from an idealized
$\beta$-Cristobalite (C9) crystal \cite{c9nrl}.
Silicon atoms always occupy the center of a perfect tetrahedron formed by 
the closest oxygen atoms above and below. In a real experiment, a thin
silica surface is effectively infinite on a microscopic scale. The positions of
the atoms in the semi-infinite silica surface region can be obtained through
periodic boundary conditions in the $xy$ directions and through extension
of the tetrahedron structure in the negative $z$ direction. 
The length of the $x$ and $y$ directions of the planar surface used in
the simulation is $L_x=45.605$ {\rm \AA} and $L_y=43.883$ {\rm \AA}
respectively and the number of atoms in each layer of the surface
is 90, 90, 270, and 90.
The top oxygen atoms form a close-packed FCC (111) structure with the
closest O-O distance of
$2a=4 \sqrt {2/3}\, b$, where $b=1.5515$\AA$\,$ is the Si-O bond length.

We first focus on the neutral silica surface. The Lennard-Jones (LJ) parameters of Si 
and O atoms in our silica model were taken from the work
of Rossky and coworkers \cite{Lee_Rossky1994,Giovambattista_Rossky2006}. 
(The hydroxylated silica surface also has partial charges on the three atoms of the H-O-Si group and 
will be discussed later.) We used the  six-site acetonitrile model of Nikitin and
Lyubartsev where intermolecular interactions from each atomic site are represented by a combination
Lennard-Jones and electrostatic potentials \cite{Nikitin_Lyubartsev2007}.
The usual combination rules,
$\sigma_{12}=(\sigma_1 + \sigma_2)/2$ and
$\epsilon_{12}=\sqrt{\epsilon_1\epsilon_2}$ are used in the calculation of mixed LJ interactions.
By treating the semi-infinite silica region from
the $n$-th layer to negative infinity as a continuum, we obtain
an integrated LJ 9-3 potential: 
\begin{equation}
\phi_c(n,z)=\sum_{i=n}^{n+3} \frac{\rho_i
\sigma_i^3}{2\sqrt{2/3}a}\frac{2\pi\epsilon_i}{3} \left[\frac{2}{15}\left(
\frac{\sigma_{i}}{z-z_i}\right)^9 - \left(\frac{\sigma_{i}}{z-z_i}\right)^3
\right] \label{eq:phic}
\end{equation}
where $\rho_i$ is the number density of atoms per unit area for the $i$-th
layer, $\sigma_i$ and
$\epsilon_i$ are the usual LJ 12-6 parameters for the type of atom in the
$i$-th layer, and $z_i$ is the location of the
$i$-th layer in the $z$ direction. 

\begin{figure}[tdp]
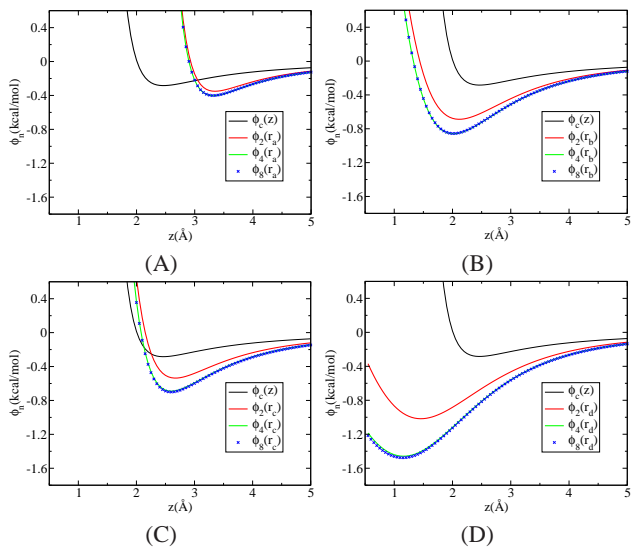

\centerline{
\begin{tabular}{cc}
\psfig{file=./newfigs/potsilica1.ps,width=1.6in,angle=270} &
\psfig{file=./newfigs/potsilica2.ps,width=1.6in,angle=270} \\
(A) & (B) \\
\psfig{file=./newfigs/potsilica3.ps,width=1.6in,angle=270} &
\psfig{file=./newfigs/potsilica4.ps,width=1.6in,angle=270} \\
(C) & (D) 
\end{tabular}} 
\caption{The potential of a test LJ atom with parameters of the
N in the acetonitrile model placed at four different typical
sites ($r_a$, $r_b$, $r_c$ and $r_d$) on the silica surface. (A) $r_a$ is
on the top of an oxygen atom.
(B) $r_b$ is on the top of the middle of two closest oxygen atoms. (C)
$r_c$ is in the middle of the corresponding $r_a$ and $r_b$. (D) $r_d$ is on
the top of the center of an equilateral triangle formed by three closest
oxygen atoms. Label for the black line is $\phi_c(z)=\phi_c(n=1,z)$. See
Eq.~\eqref{eq:phin}}
\label{fig:potsilica}
\end{figure}

The total VdW potential at
position ${\mathbf r}$ from the whole silica surface including both $n$
layers of explicit atoms and the semi-infinite continuum region is:
\begin{multline}
\phi_n({\mathbf r}) =  \sum_{i=1}^n \sum_{j=1}^{N_i} 4\epsilon_i \left[
\left(\frac{\sigma_i}{|{\mathbf r}-{\mathbf r}_j^i|} \right)^{12} - 
\left(\frac{\sigma_i}{|{\mathbf r}-{\mathbf r}_j^i|}\right)^6 \right]\\ +
\phi_c(n+1,z)
\label{eq:phin}
\end{multline}
where $N_i$ is the number of explicit atoms of the $i$-th layer (including periodic images up to a specified cutoff distance) and
${\mathbf r}_j^i$ is the position of the $j$-th atom of the $i$-th layer.
A plot of $\phi_n({\mathbf r})$ with $n$ = 0, 2, 4, and 8 respectively for a test
LJ atom with the LJ parameters of N in the acetonitrile model as a function of its
distance from the silica surface is shown in Fig.~\ref{fig:potsilica}.
 
Figure \ref{fig:potsilica} shows that the potential calculated using four
explicit layers overlaps that of eight explicit layers for all positions
tested.
Clearly, using four explicit layers with a continuum potential is accurate
enough to reproduce the VdW potential of the semi-infinite region. Note that
noncharged atoms prefer to
dock into the position ${\mathbf r}_d$ rather than the other three because of
its lower energy minimum.

\begin{figure}[tbp]
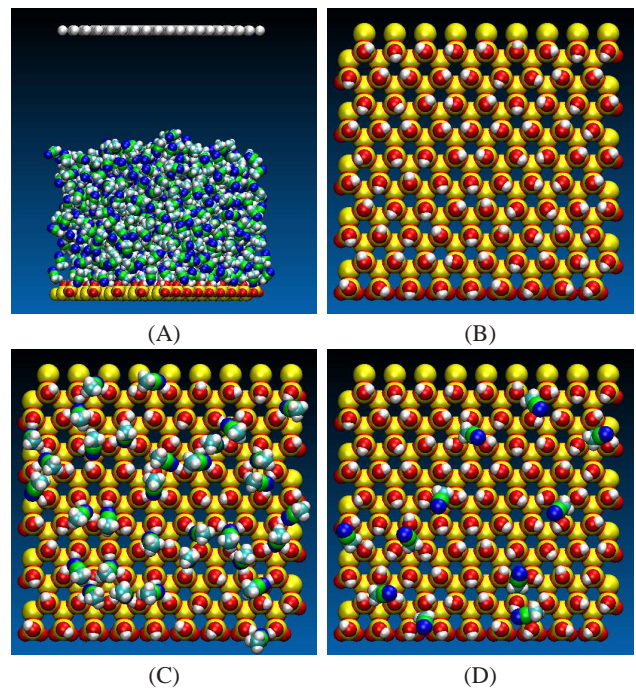

\centerline{
\begin{tabular}{cc}
\psfig{file=./newfigs/simulationset.eps,width=1.6in,angle=0} &
\psfig{file=./newfigs/silicasurface.eps,width=1.6in,angle=0}  \\
(A) & (B) \\
\psfig{file=./newfigs/dockingN.eps,width=1.6in,angle=0} &
\psfig{file=./newfigs/dockingH.eps,width=1.6in,angle=0} \\
(C) & (D) 
\end{tabular}}
\caption{Side view of the simulation
setup (A), with a top view of the hydroxylated silica surface (B). 
Top views of the CN (C) and CH$_3$ groups (D) of the first sublayer of acetonitrile
molecules docking into the hydroxylated silica surface in a
typical configuration  as visualized by
VMD software \cite{vmd1996} are shown.}
\label{fig:silica}
\end{figure}

The hydrogens in the hydroxylated silica surface used in this work are all
attached to the oxygen atoms of the top surface layer (see
Fig.~\ref{fig:silica} (B)).
The entire silica
surface (all Si atoms and O atoms) is fixed, with the top oxygen layer at the
$z=0$ plane, except that hydrogens can freely rotate. 
The length of O-H bond is constrained to be $1.0 {\rm \AA}$ and the Si-O-H
harmonic angular bending
potential has an equilibrium angle of $109.27$ degrees and a large
bending strength of $200.0 {\rm kcal/mol}$.

\begin{figure}[tdp]
\centerline{
\psfig{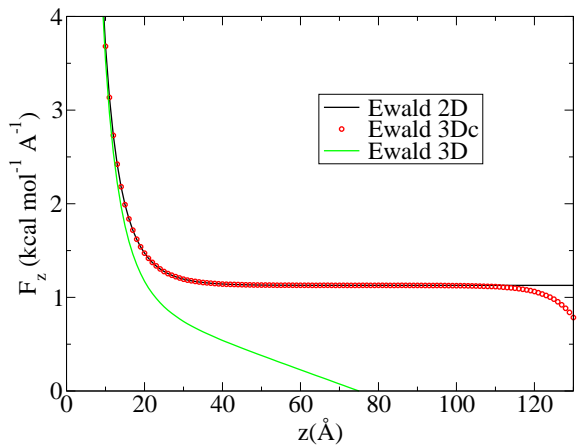} }
\caption{The electrostatic interaction between two unit point charges computed
by three different methods. The simulation setup has a length scale of
$l_z=150$ {\rm \AA} and $l_x=l_y=43$ {\rm \AA}.This
test case is similar to Figure 4 in reference \cite{Yeh_Berkowitz1999}
but for the specific length scales used in our simulation.}
\label{fig:cmp2d3dewald}
\end{figure}

The hydroxylated silica surface has partial charges on the three atoms of
the H-O-Si group, which are also taken from the work of Rossky and
coworkers \cite{Lee_Rossky1994,Giovambattista_Rossky2006}. The long-ranged
electrostatic interaction has to be treated carefully.
We have carried out a test
on a pair of ions in a slab-like box with the same length scales used in our simulation
of the acetonitrile system (see below) using the exact
Ewald2D method \cite{Heyes_Clarke1977}, the Ewald3D method, and the
corrected Ewald3D method introduced by Yeh and Berkowitz
\cite{Yeh_Berkowitz1999}.
Fig.~\ref{fig:cmp2d3dewald} shows a comparison of these three methods for
the test case. Clearly the standard Ewald3D method is very inaccurate but the corrected Ewald3D is
in close agreement with the Ewald2D method for separations less than about
$100 {\rm \AA}$.
In what follows we use the corrected
Ewald3D method, which is less computationally expensive but quite
accurate in the region from $z=0$ to $75 {\rm \AA}$.

\begin{figure*}[htdp]
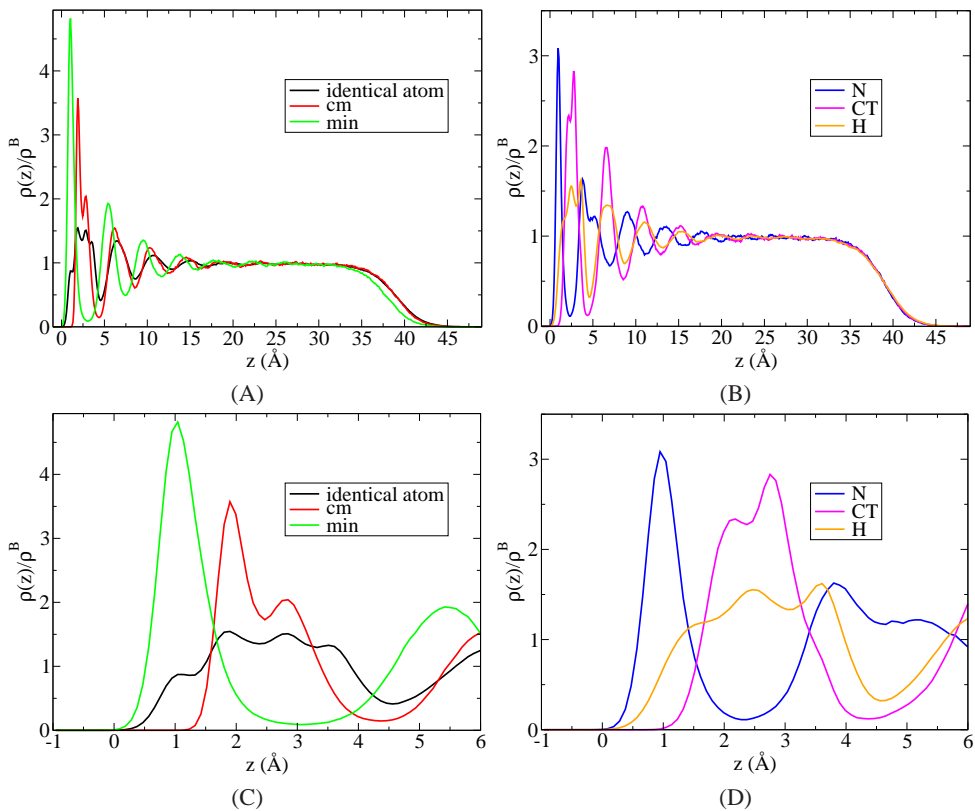

\begin{tabular}{cc}
\psfig{file=./newfigs/allcmminzden-onSilica.ps,width=2.5in,angle=270} &
\psfig{file=./newfigs/YN-CT-HonSilica.ps,width=2.5in,angle=270} \\
(A) & (B) \\
\psfig{file=./newfigs/closer-allcmminzden-onSilica.ps,width=2.5in,angle=270} &
\psfig{file=./newfigs/closer-YN-CT-HonSilica.ps,width=2.5in,angle=270} \\
(C) & (D)
\end{tabular}
\caption{Density profiles of the interfacial system defined from positions
of all atoms treated as identical, positions of the center of mass,
and the minimum position of acetonitrile molecules (A) and positions of the nitrogen
atoms, alkyl carbon atoms, and hydrogen atoms of acetonitrile (B). The existence
of three qualitatively different regions is evident: a LS region near the silica surface
strongly perturbed by the substrate, a bulk liquid-like region from about 22 to 33\AA\
 and a LV interface at larger $z$. C and D give an expanded 
view  of A and B respectively from the surface to 6\AA\ . The density
is normalized by the bulk value $\rho^B$ from a separate bulk
simulation. The minimum position of an acetonitrile molecules refers to the
minimum z position of all six atoms of the molecule. T=298K and CT stands
for the alkyl carbon atom.}
\label{fig:denprofile} 
\end{figure*}

\subsection{Acetonitrile model and simulation parameters}
We performed molecular dynamics simulations of acetonitrile
confined in the $z$-direction by two parallel walls using the software DL\_POLY version
2.18 \cite{dlpoly} with modifications to include external potentials and
the corrected Ewald3D method to treat the electrostatics. 
The acetonitrile system studied here consisted of 864 molecules. 
All acetonitrile molecules were confined by a hydrophobic wall located at
$75$\AA$\,$ and the silica wall with its top oxygen layer located at
$0$\AA. The potential of the hydrophobic wall had the same form as in
Eq.~\eqref{eq:phic} but uses only the repulsive
part of the potential \cite{Weeks_Chandler_Andersen1971}.
Periodic boundary conditions were employed with 
$L_x=45.605$\AA, $L_y=43.883$\AA, and $L_z=150$\AA.
The empty region between $75 {\rm \AA}$ and $150 {\rm \AA}$ is required
for the application of the corrected Ewald3D method (see
Fig.~\ref{fig:cmp2d3dewald}).

The simulation setup is shown in Fig.~\ref{fig:silica}(B).  The
acetonitrile potential energy parameters were those previously published by Nikitin and
Lyubartsev \cite{Nikitin_Lyubartsev2007}. The cutoff distance for the VdW
and the short-ranged part of the corrected Ewald simulation was 15{\rm \AA}.
The screening parameter for the
Ewald summation was 0.26${\rm \AA}^{-1}$. The number of $k$-space vectors 
for the Ewald summation was 15, 15, and 45 for the $x$, $y$, and $z$
directions respectively. These
parameters were optimized choices yielding the best efficiency for our system
setup.

A bulk simulation was initially
equilibrated at T = 298K and P = 1 atm before putting the acetonitrile on the
silica surface.
The interfacial system was equilibrated for nearly a nanosecond
at constant number of molecules, constant volume and
constant temperature T=298K (NVT ensemble) using the Berendsen
method \cite{Berendsen1984} until no further energy drift was observed. To
reduce the computational cost, we first employed a more efficient but less
accurate method \cite{Wolf1999,Chen_Weeks2004} with truncated Coulomb interactions
for several hundred picoseconds and
then applied the accurate corrected Ewald3D method in the later stage of
the equilibration. The latter was also
used in all production runs. One NVE
production simulation was run for a duration of 300ps using a time step of
1 fs. This was followed by another NVE simulation of 50ps using a time step of 0.5
fs to permit the accurate determination of time correlation functions at very short times. Positions
of atoms were recorded at every 5 fs and every 1 fs
respectively for further analysis. Total energy is well conserved and no
temperature drift was observed in the NVE production simulation. 

A separate simulation of the bulk acetonitrile system was carried out for comparison. The
production simulation of the bulk system had 864 molecules in a
cubic box of size 42.37\AA. The number density was $\rho^B=0.774{\rm
g/cm^3}$, which is in good agreement with the experimental value of
$\rho=0.777{\rm g/cm^3}$ at the temperature T=298K \cite{Gallant1969}. 
A top view of the surface and the first sublayer acetonitrile molecules on the
surface is shown in Fig.~\ref{fig:silica} (B), (C) and (D).

\section{Results and Discussions}
\label{sec:results}
\subsection{Singlet densities and orientational profiles}

Stable LV and LS interfaces are formed in the
simulation of  the acetonitrile system and are most simply
described by the singlet density profile.
However with larger molecules like acetonitrile there are several
reasonable choices for which atomic type or molecular position should be used to
characterize the location of the molecule in the density profile.

In Fig.~\ref{fig:denprofile} we show density profiles defined from positions of the
nitrogen (N), methyl carbon (CT), hydrogen (H), center of mass (cm),
and minimum $z$ position of any atom in the molecule. We also defined an
``identical atom'' profile in which all atoms in the molecule are treated as if they were
of the same type and we report the spatial distribution of this single atomic species.
The average of the number density $\rho(z)$ over the region $25 {\rm \AA} <
z < 30 {\rm \AA}$ ($0.98 \rho^B$) is nearly the same as the bulk density using all
definitions and we consider this the bulk region of our interfacial system.

Following previous work \cite{Paul_Chandra2005}, the LV
interfacial region is defined as
the region over which the number density decreases from 90\% to 10\% of the
bulk density. As shown in Fig.~\ref{fig:denprofile}
(A) and (B), density profiles from all definitions except the minimum
position yield almost identical LV interfacial regions ($35.24{\rm \AA} < z < 42.08{\rm \AA}$),
while the minimum position shifts the LV interfacial region to
$34.20 {\rm \AA} < z < 40.94 {\rm \AA}$. The thickness of the LV interface is about $6.7
{\rm \AA}$,  which is slightly larger than the value $4.7{\rm \AA}$ reported
by Paul and Chandra \cite{Paul_Chandra2005} using a three-site model of
acetonitrile.

As would be expected, the LS density profiles are much more
structured than those of the LV interface and have different forms depending
on the choice of defining molecular position. Clearly, acetonitrile
forms layers on the silica surface from $z=0$ to about $z\simeq 20{\rm \AA}$ and the
various profiles provide information about different features of this layering.
It is interesting that the indications of the surface-induced layering persist past 20 \AA\ 
from the substrate. This contrasts with water near silica where perturbations die off by about 10 \AA\
\cite{Lee_Rossky1994}, probably indicating the greater 
orientational flexibility of the tetrahedral hydrogen-bond network.

Most density
profiles show shoulder peaks except the
density profile based on the minimum position. The first layer of
acetonitrile adsorbed on silica is defined as the region from $z=0$ to the
first major depletion of density ($z=3 {\rm \AA}$ for the minimum position and
$z\simeq 4.5{\rm \AA}$ for the identical atom profile and the center of mass
profile). The first minimum of the density profile of the
nitrogen atom appears at a $z$ distance
less than $3{\rm \AA}$ (Fig.~\ref{fig:denprofile} (D)), indicating that molecules with
the nitrogen atom pointing to the silica
surface likely dominate at close distances to the surface.

We further divide the first layer into two sublayers according to the
shoulder peak position of the density profile of identical atoms $z=1.2{\rm \AA}$
(Fig.~\ref{fig:denprofile} (C)). A top view of the molecules within the
sublayer $0<z<1.2{\rm \AA}$ was shown in Fig.~\ref{fig:silica} (C)
and (D). Clearly molecules in the first sublayer prefer to orient with their
nitrogen atoms binding to hydrogen atoms (Fig.~\ref{fig:silica} (C)).
About 25\% of the sublayer molecules dock into the equilateral
triangle formed by three closest oxygen atoms that have their attached
hydrogen atoms pointing outside the triangle (Fig.~\ref{fig:silica} (D)).

\begin{figure}[tdp]
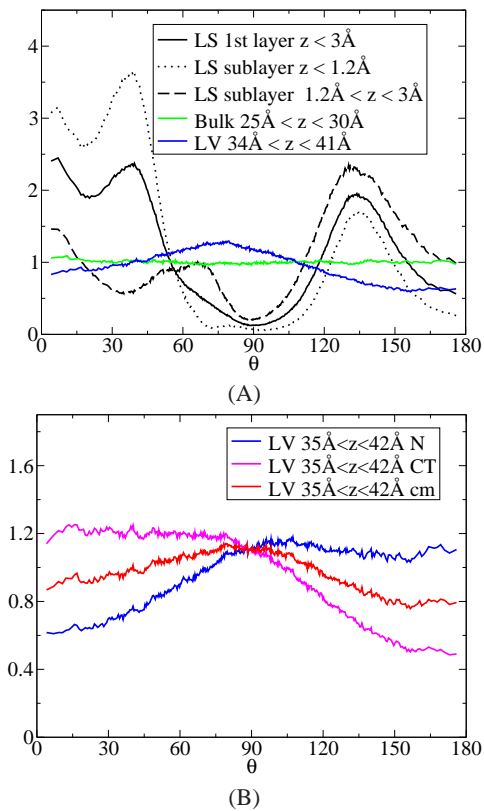

\begin{tabular}{c}
\psfig{file=./newfigs/angledistri-LS-LV-Bulk.ps,width=2.5in,angle=270} \\
(A) \\
\psfig{file=./newfigs/LV-orientationdiff.ps,width=2.5in,angle=270} \\
(B) 
\end{tabular}
\caption{Orientational profiles of the acetonitrile dipoles relative to the
surface normal (positive $z$ direction) when $z$ is the minimum position of
the acetonitrile molecules (A) and the orientational profiles of the LV
interface when $z$ is the position of nitrogen atom,
methyl carbon atom and center of mass (B). The Jacobian factor $\sin\theta$ has been
scaled out in all plots of angle distributions, so the bulk system has a uniform distribution.
$\theta$ is the angle between the dipole orientation (CT $\leftarrow$ N)
and the positive $z$ direction.
In our geometry $\theta=0$ implies that the CN group points to the liquid
phase at the LV interface and to the silica surface
at the LS interface.}
\label{fig:thetaall}
\end{figure}

Once we have defined the sublayer, the first layer, the bulk region, and the
liquid-vapor region according to singlet density profiles, it seems natural to
examine the orientational distributions in these regions as shown in
Fig.~\ref{fig:thetaall} (A). A consistent picture of the LS interface
arises from all profiles. The orientation of an acetonitrile molecule in
the first layer of the LS region tends to lie in two branches: $\theta
< 45^{\circ}$ where the nitrogen end points to the
surface and around $\theta\simeq 130^{\circ}$ where the nitrogen end points away
from the surface. Most molecules within the sublayer $z< 1.2$ {\rm \AA} lie
in the first branch
$\theta < 45^{\circ}$ while more molecules in the sublayer $1.2$ {\rm \AA}$ <
z < 3$ {\rm \AA} lie in the second branch so the structure is reminiscent of a lipid
bilayer. As the distance between
an acetonitrile molecule and the silica surface increases, the
orientational distribution gradually approaches
unity, where the dipole of a molecule orients randomly as in the bulk. 

However, results in the LV interfacial region appear to depend on
the choice of singlet density profile. When the location of a molecule is defined
by its minimum position as in Fig.~\ref{fig:thetaall} (A), the orientational profile
has a small maximum at $\theta=90^{\circ}$, parallel to the interface.
But Fig.~\ref{fig:thetaall} (B) shows that the orientation profile of molecules
defined using the positions of N, CT and cm seem to give different
interpretations of the preferred molecular orientation at the LV
interface. When the position of the center of mass is used, more molecules
seem to orient parallel to the interface, consistent with results from the use
of the minimum position of molecules in Fig.~\ref{fig:thetaall}
(A). However, as noted in \cite{Cheng_Berne2009}, more CH$_3$ groups of the 
acetonitrile molecules appear to point into the liquid (gas) phase
if we define the location of an acetonitrile molecule using the position of
its nitrogen (carbon) atoms.

We show here that this apparent discrepancy arises from the different
contributions of ``over-counted'' dipoles, arising from the different definitions of molecular position. These considerations become important only when the size of the molecule is a finite fraction of the interface width, and when they are taken into account
results from the different definitions can be related to each other.

A molecular $+/-$ dipole is defined as  ``over-counted'' when
its positive side (CT atom in the case of acetonitrile) falls into the region
of interest ($35$ {\rm \AA}$ < z < 42$ {\rm \AA} for the LV interface), while its negative side (the N atom in this case) is still located outside of the region, with a
similar definition
for an over-counted $-/+$ dipole with the negative side inside and positive side outside. 
The difference shown in
Fig.~\ref{fig:thetaall} (B) using definitions from the  CT atom and N atom can
be expressed as $g_{\rm over,+/-}(\theta) - g_{\rm over,-/+}(\theta)$. 

The over-counted contribution of $+/-$ dipoles can also be related to the density and angular distribution of the dipoles based on the profile defined using the center of mass position:
\begin{eqnarray}
g_{\rm over,+/-}(\theta) =\left\{ \begin{array}{c} 
\frac{\displaystyle \int_{z_1-L/2\cos\theta}^{z_1+L/2\cos\theta}dz\,
\rho_0(z)f_0(z,\theta)}{\displaystyle \left|\int_{z_1}^{z_2}dz\,
\rho_+(z)\right| }  \\ \\
\quad\quad\quad\quad{\rm for} \quad 0 \leqslant \theta < \pi/2 \\ 
\\
\frac{\displaystyle \int_{z_2+L/2\cos\theta}^{z_2-L/2\cos\theta}dz\,
\rho_0(z)f_0(z,\theta)}{\displaystyle \left| 
\int_{z_1}^{z_2}dz\, \rho_+(z) \right| }  \\ \\
\quad\quad\quad\quad{\rm for} \quad \pi/2 
\leqslant \theta \leqslant \pi \end{array} \right.
\label{eq:overcountg}
 \end{eqnarray}
where $L$ is the length of the dipole (distance between CT and N).
Here $\rho_0(z)$ is the density profile defined using the center of mass position
and $\rho_+(z)$ is the profile when the location of the molecule is
defined according to its atom with positive charge (these functions closely resemble each
other and either can be used in the calculation).
$f_0(z,\theta)d\theta$ is the probability for
a given dipole with its center located at position $z$ having
orientational angle in $[\theta,\theta+d\theta]$.
When $+/-$ and $z_1/z_2$ are exchanged simultaneously, Eq.~(\ref{eq:overcountg})
then gives the contribution of over-counted $-/+$ dipoles.

\begin{figure}[tdp]
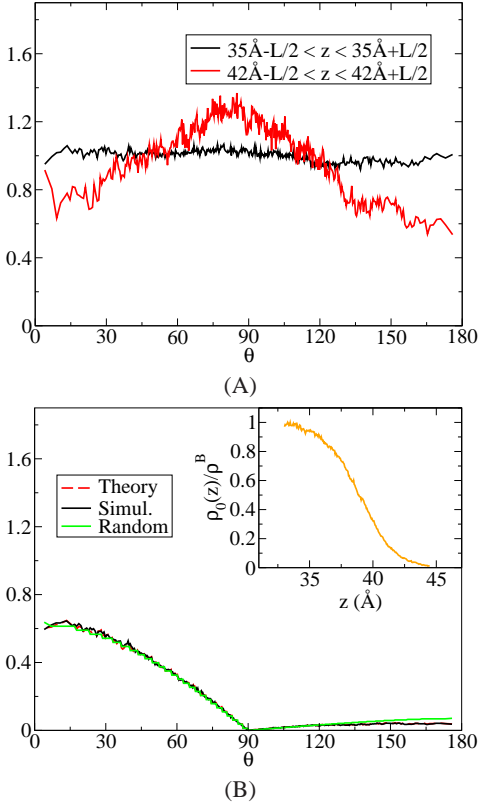

\begin{tabular}{c}
\psfig{file=./newfigs/middle-around35A-42A.ps,width=2.5in,angle=270}\\
(A) \\
\psfig{file=./newfigs/numerical-simulcmpLVoverangle.ps,width=2.5in,angle=270} \\
(B) \\
\end{tabular}
\caption{Orientational profile of molecules at the boundaries
around $z_1=35{\rm \AA}$ and $z_2=42{\rm \AA}$ (A) and
contribution from the over-counted $+/-$ molecules to the
orientational distribution (B). 
The black solid line
(Simul.) in (B) is computed from the histogram of the over-counted molecules.
The red dashed line (Theory) in (B) is computed through Eq.~\eqref{eq:overcountg} using 
$\rho_+(z)\simeq\rho_0(z)$, shown
in the inset of (B) (see also Fig.~\ref{fig:denprofile}), and
$f_0(z,\theta)$ shown in (B) with $z_1=35{\rm \AA}$, $z_2=42{\rm
\AA}$ and $L=2.6 {\rm \AA}$. The green solid line (Random) in (B) is computed in the
same way as the red dashed line except that $f_0(z,\theta)=1$.}
\label{fig:overcountg}
\end{figure}

In order to check the validity of Eq.~\eqref{eq:overcountg},
we first calculate the average orientational distribution around $z=z_1$ and
$z=z_2$ as shown in Fig.~\ref{fig:overcountg} (A).
Figure \ref{fig:overcountg}(B) shows the comparison between the calculation from
Eq.~\eqref{eq:overcountg} (red dashed line) when the average orientational
distribution is used as
the input and when the explicit histogram distribution of the overcounted
molecules is used (solid black line). The overlap between the two lines confirms the validity of
Eq.~\eqref{eq:overcountg}.
In fact, Eq.~\eqref{eq:overcountg} can be accurately
approximated in this case by simply taking $f_0(z,\theta)=1$. The computed
over-counted contributions are barely changed. As shown in Fig.~\ref{fig:overcountg}(B),
Eq.~\eqref{eq:overcountg} gives an excellent description of the
simulation results.

\begin{figure*}[tdp]
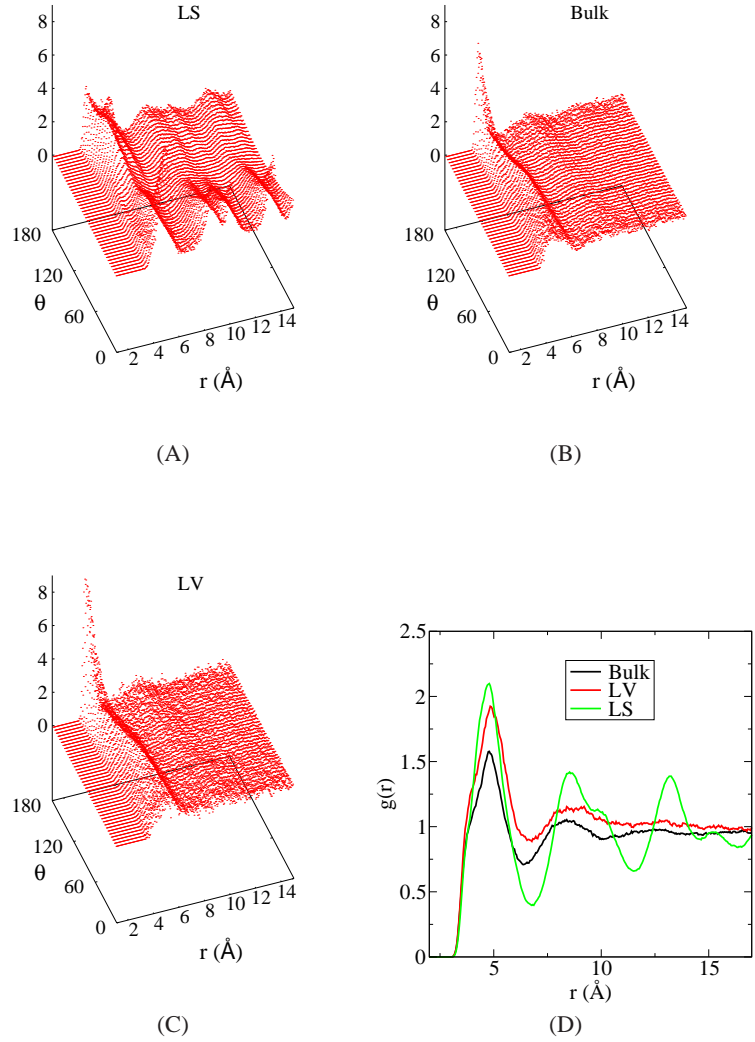

\centerline{
\begin{tabular}{cc}
\psfig{file=./newfigs/viewinter6-34LSilica.eps,width=2.0in,angle=0} &
\psfig{file=./newfigs/viewinter6-34Bulk-25-30A.eps,width=2.0in,angle=0} \\
(A) & (B)  \\
\psfig{file=./newfigs/viewinter6-34LV.eps,width=2.0in,angle=0} &
\psfig{file=./newfigs/cmgrLVS298K.ps,width=2.0in,angle=0}  \\
(C) & (D) \\
\end{tabular}}
\caption{Radial angular distribution $g_{cm}(r,\theta)$
at T = 298K for the first layer LS region (A), the bulk liquid region (B),  
the interfacial LV region (C),
and the results after angular integration over $\theta$ (D). The 
center of mass position is used for the definition of three different regions. 
 $\theta$ is the angle between the dipole axis of two
molecules and $r$ is the distance between the center of mass positions of
the two molecules.}
\label{fig:viewbulkinter}
\end{figure*}

Clearly, because the number of over-counted molecules at the high density side $z=z_1$
of the LV interface cannot be compensated by those at the
low density side $z=z_2$ and the magnitude of $g_{\rm over,+/-}(\theta)$ is non-negligible,  
the orientational profiles computed from different
definitions of the position of a molecule in the interface give
apparently contradictory predictions for the prefered orientation of the
molecules.

In the case of a bulk system,
the over-counting from the two boundaries exactly cancels each other and
$g_{\rm over,+/-}(\theta) = g_{\rm over,-/+}(\theta)$. In the case of the
LS interface, because values of $\rho_0(z)$ in the numerator of
Eq.~\eqref{eq:overcountg} is much smaller 
than the average density of the interfacial region (e.g. $\rho_0(z=4.5
{\rm \AA}) \simeq 0 )$, the magnitude of $g_{\rm over,+/-}$ is very small
and makes almost no contribution to the overall orientational profile in the
interfacial region (e.g. the first layer orientational profile). Therefore,
there is little discrepancy for cases of bulk and solid-liquid interfacial
systems. 

In case of the LV interface, when the length of the dipole
(e.g. $L=2.6 {\rm \AA}$ for an acetonitrile molecule) is comparable to the
interfacial width (e.g. $z_2-z_1\simeq 7{\rm \AA})$, $g_{\rm over,-/+}$ is
asymmetric and the over-counted molecules from the dense boundaries can make
a significant contributions to the overall distribution and thus generate
an apparent discrepancy. These considerations should be taken into account in
interpretations of simulation or experimental data. 

\subsection{Correlated structural properties}

Although singlet density and orientational profile have been studied in previous 
work \cite{Morales_Thompson2009,Paul_Chandra2005}, correlated
structures are seldom investigated. In order to shed light on how the
antiparallel correlation \cite{Nikitin_Lyubartsev2007} between two
acetonitrile molecule changes due to the effect of inhomogeneity, we
define a quantity called the radial angular distribution function:
\begin{equation}
g_{cm}(r,\theta) = \frac{1}{N_c}\sum_{i=1} ^{N_c}
\sideset{}{'}\sum_{j=1}^{N_c}\left<
\delta(r-r_{ij}) \delta(\theta-\theta_{ij}) \right> ,
\end{equation}
where $N_c$ is the number of
particles contained in the three different regions defined above: the bulk region, LV region,
and the first layer of the LS region. The $g_{cm}(r,\theta)$ is essentially
the normalized histogram over independent pairs of molecular dipoles in the
three regions of interest; the prime on the sum indicates that terms with $i=j$ are excluded.
Here $r_{ij}$ is the distance between the
center of mass of the $i$-th
and $j$-th molecule and $\theta_{ij}$ is the angle between the dipole
orientations of the two molecules. The radial angular correlation function
is normalized to $1$ at large distance $r$ and the integral over
$\theta$ gives the usual center of mass radial distribution function.

Fig.~\ref{fig:viewbulkinter} shows that acetonitrile at the LV
interface has an even stronger
antiparallel correlation than in the bulk. In contrast, molecules in the first layer of the
LS interfacial region completely lose the preferred correlation at
$\theta=180^\circ$.
The silica surface not only aligns the singlet molecular orientation such
that it prefers to point toward the normal of the surface
(Fig.~\ref{fig:thetaall}(A)),
but it also significantly destroys the strong anti-parallel
pair correlations seen for bulk liquid acetonitrile. At the free liquid-vapor interface, the singlet
orientation differs only slightly from that of the random distribution in
the bulk. The dominant intermolecular interactions
cause LV interfacial molecules to preferentially orient in an
antiparallel manner. But local interactions from the silica surface are strong
enough to significantly
modify the pair structure of molecules in the first layer.
Since solid acetonitrile has even stronger anti-parallel correlations
that would be similarly disrupted at the silica surface this, along with
the general modification of bulk structure at a surface, could
provide a possible explanation for the decrease of
the melting point when acetonitrile is confined in silica pores.

\subsection{Singlet and collective rotational dynamics}
\begin{figure}[tdp]
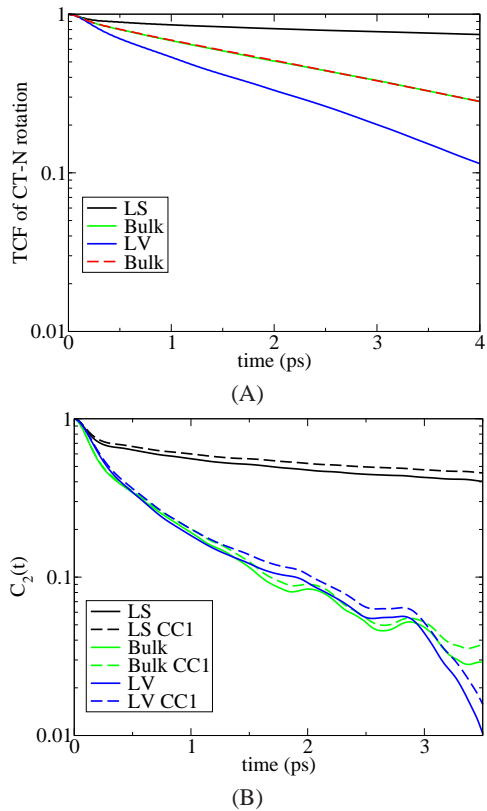

\centerline{
\begin{tabular}{c}
\psfig{file=./newfigs/singlerotation.ps,width=2.5in,angle=270} \\ (A) \\
\psfig{file=./newfigs/collectR-allorder-1storder.ps,width=2.5in,angle=270} \\
 (B) \\
\end{tabular}}
\caption{Singlet (A) and collective (B) reorientational correlations for
acetonitrile in different regions. See Eqs.~\eqref{eq:tcfct} and
~\eqref{eq:tcfc2}. The red dashed line in (A) is computed from
a separate simulation of the bulk system. Dashed lines in (B) are the
corresponding CC1 model calculations.  Rotational time scales obtained by linear
fit of the logarithm plot after $t> 2{\rm ps}$ are shown in
Table~\ref{tab:rscales}. The definition of different regions is the same as
in Fig.~\ref{fig:thetaall}.}
\label{fig:rotations}
\end{figure}
Both singlet and collective rotational dynamics are of interest as they can
be measured from spectroscopic
experiments \cite{Loughnane_Fourkas1999,Kittaka2007}. The time correlation
function (TCF) of the rotation of the symmetrical axis (CT-N) is defined as
\begin{equation}
C(t) = \frac{1}{N_c}\sum_{i=1}^{N_c} \left< {\mathbf v}_i(0) \cdot 
{\mathbf v}_i(t) \right> ,
\label{eq:tcfct}
\end{equation}
where ${\mathbf v}_i$ is the rotational axis of the $i$-th molecule and
$N_c$ is the total number of molecules counted
in particular region of interest. The
collective reorientational behavior of the bulk acetonitrile system has been
studied by Ladanyi and coworkers \cite{Elola_Ladanyi2005} but few workers have
addressed the collective dynamics of interfacial or confined systems.
  
The collective polarizability $\overline{\Pi}$ of a group of molecules is
the sum of polarizabilities of each molecule. The standard point
dipole/induced point dipole (DID) model expresses the
individual polarizability $\overline{\pi}(j)$ for molecule $j$ as a sum of a molecular
component (gas-phase polarizability $\overline{\alpha}(j)$) and an interaction-induced component
\begin{equation}
\overline{\pi }(j) = \overline{\alpha}(j) + \overline{\alpha}(j)\cdot
\sum_{k\not=j}^{N} \overline{T}_{jk}\cdot\overline{\pi}(k),
\label{eq:did}
\end{equation}
where the dipole-dipole tensor $\overline{T}_{jk}$
between molecule $j$ and $k$ can be written as
\begin{equation}
 \overline{T}_{jk} = \frac{1}{r^3}\left.(3\hat{r}\hat{r} -
\overline{\rm I})\right|_{{\mathbf r}={\mathbf r}_{jk}}
\end{equation}
and $\hat{r}={\mathbf r}/r$ is the unit vector along the direction
${\mathbf r}$. Eq.~\eqref{eq:did} can be solved self-consistently or
by matrix inversion \cite{Applequist1972}.
An efficient first-order approach (CC1) corresponds to
replacing $\overline{\pi}(k)$ on the right hand side of
Eq,~\eqref{eq:did} by the gas-phase polarizability
$\overline{\alpha}(k)$ \cite{Elola_Ladanyi2005}. 

The total collective
polarizability $\overline{\Pi}$ for a region of $N_c$ molecules is 
\begin{equation}
\overline{\Pi} =\sum_{j}^{N_c}\overline{\pi}(j)= \overline{\Pi}_0
\overline{\rm I} +
\overline{\Pi}_2  ,
\end{equation}
where $\overline{\Pi}_0$ and $\overline{\Pi}_2$ are the isotropic and
anisotropic polarizability of the region of $N_c$ molecules respectively.
Different from the bulk system, the average of the $xz$ or $yz$
component of the anisotropic polarizability in the nonuniform system we study here
is not zero. In order to focus on just the relaxation part of the anisotropic
collective polarizability, we define the TCF $C_2(t)$ as
\begin{equation}
C_2(t) = \left< {\rm
Tr}[(\overline{\Pi}_2(0)-\left<\overline{\Pi}_2\right>)\cdot 
(\overline{\Pi}_2(t)- \left<\overline{\Pi}_2\right>) ] \right>
\label{eq:tcfc2}
 \end{equation}
Analysis of $C_2(t)$ is crucial because it is directly relevant to the OKE
signal \cite{McMorrow_Kenney-Wallace1988,Geiger_Ladanyi1989,Hu_Margulis2008}.
The notation used in this subsection is the same as the work by Hu {\it et. al.}
\cite{Hu_Margulis2008}. The TCFs of anisotropic collective polarizability
for different regions of our system are primarily studied here using both
the all-order and first order DID model. The gas phase polarizabilities of
acetonitrile used here are $\alpha_{\parallel}=5.80$ {\rm \AA}$^3$ and
$\alpha_{\perp}=3.65$ {\rm \AA}$^3$ \cite{Geiger_Ladanyi1989}. The
polarization effect from the atoms of
the silica surface is not considered here.

\begin{table}[tdp]
\center{\caption[]{\label{tab:rscales} Estimate of singlet and collective rotational
time scales for acetonitrile in different regions.\footnotemark[1]}}
\begin{tabular}{cccc} \hline
axis/model & LS 1st layer & Bulk region & LV region \\ \hline
CT-N  &  24(4) ps    &  3.4(2) ps     & 1.9(1) ps  \\
Collective   &  8.1(2) ps     &  1.4(1) ps     &  1.1(1) ps  \\
CC1   &  9.6(2) ps     &  1.5(1) ps     &  1.2(1) ps  \\   
\hline
\end{tabular}
\footnotetext[1]{Numbers in parentheses are uncertainties in the final digit.
Random errors from the least-square fits of the data from 2ps to 4ps
(Fig.~\ref{fig:rotations} A)  or 2ps to 3.5ps (Fig.~\ref{fig:rotations} B)
are much smaller than systematic errors that likely arising from this limited fitting
range. The latter are crudely estimated here based on small variations of input data.
}
\end{table}
Fig.~\ref{fig:rotations} shows the singlet and collective reorientation
correlations in a logarithmic plot.  A separate
calcuation of $C(t)$ from the simulation of bulk acetonitrile (red dashed
line in
Fig.~\ref{fig:rotations}(A)) overlaps the corresponding $C(t)$ for the
bulk region  in the inhomogeneous system. The rotational time scale in the
region with a distance of larger than $25${\rm \AA} from the surface is
almost the same as in the bulk. This finding is consistent with the
experiment work of Kittaka and coworkers \cite{Kittaka2005}.  Comparison
between the all-order calculations and the first order approximation in
Fig.~\ref{fig:rotations} (B) shows that the effect of the higher orders is
almost negligible in the bulk and in the LV region, consistent with the
work of Elola and Ladanyi \cite{Elola_Ladanyi2005}. 

Higher order effects are a little more important
for the collective reorientation of molecules in the first layer
close to the silica surface. The values of the rotational time scales
determined by fitting the long time behavior (2-4 ps for singlet rotations and 2-3.5 ps
for collective rotations) to an exponential form
are shown in Table~\ref{tab:rscales}.  The value
of the time scale for the self-rotational correlation in the bulk and LV
region are $\tau_{\rm B}=$ 3.4ps and $\tau_{\rm LV}=$ 1.9ps respectively.
These values are different from
those reported by Paul and Chandra \cite{Paul_Chandra2005}, 10.75ps and
5.75ps respectively, because they determined the orientational relaxation time scale from the
time integral of the correlation function, and finite size and equilibration effects
would be expected to affect both these estimates in different ways.
Both work agrees however that the reorientational motion has been
significantly enhanced at the LV interface relative to the bulk phase.

Both singlet and collective reorientation of the first layer of acetonitrile
close to the silica surface are highly hindered. However,
the collective reorientation in the first layer is much faster
than the self-rotation of a molecule in the same region.
Fourkas and coworkers measured the collective reorientation time scales in
the bulk and in the pores as 1.66ps and 26ps respectively at T=290K and
1.42ps and 19ps respectively at T=309K \cite{Loughnane_Fourkas1999a}.
We estimated the experimental values for
T=298K, 1.54ps and 22 ps by simply taking the average of the values at the
higher and lower temperatures. 

Thus, our calculated bulk
value 1.39ps is about 10\% smaller than the experimental value 1.54ps.
Because experiments on acetonitrile
in pores measure components arising from both bulk-like and surface molecules, and
can be affected by other aspects of the confinement, our computed
time scale for the collective reorientation of the 
first layer LS acetonitrile, 8.1ps seems in reasonable qualitative agreement
with the reported experimental value of about 22ps. However, a quantitative understanding of the
different time scales in the OKE experiment and test of the triexponential
model used \cite{Loughnane_Fourkas1999a}
would require a similar theoretical investigation of the OKE signals in silica pores, which
is beyond the scope of this paper.

\begin{figure*}[tdp]
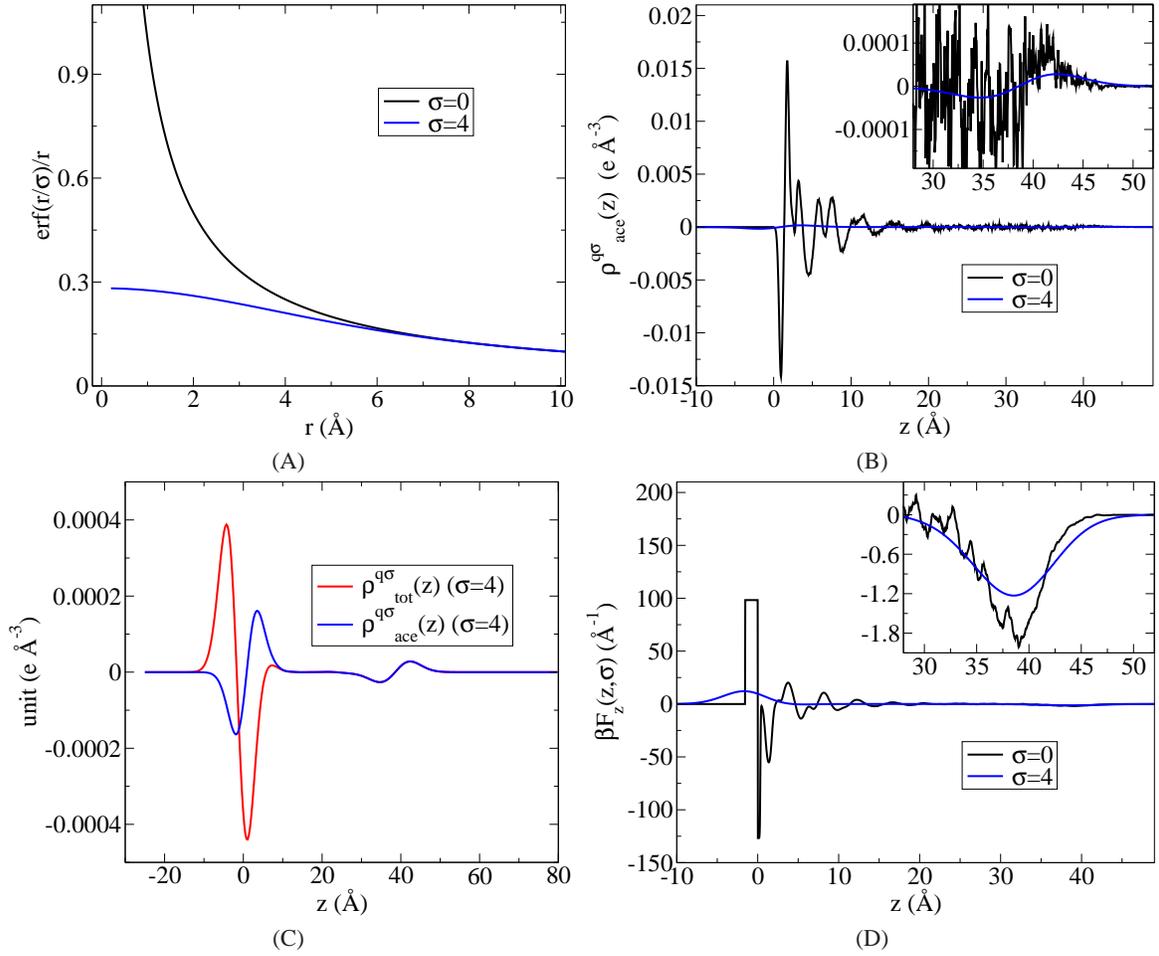

\centerline{
\begin{tabular}{cc}
\psfig{file=./newfigs/erfsep.ps,width=3.0in,angle=270} &
\psfig{file=./newfigs/acechargesep-aceLiqonSilica.ps,width=3.0in,angle=270} \\
(A) & (B) \\
\psfig{file=./newfigs/smoothchargetotace.ps,width=3.0in,angle=270}
&
\psfig{file=./newfigs/totalforcesep-aceLiqonSilica.ps,width=3.0in,angle=270}  \\
(C) & (D) \\
\end{tabular}}
\caption{Separation of the electrostatic interactions. (A) The potential from
a point charge ($\sigma =0$) and from a Gaussian charge distribution with width
$\sigma =4$. (B) The bare charge density of the mobile acetonitrile molecules and the
corresponding smooth charge density.
(C) The comparison between the smooth charge densities of the total system
and the mobile acetonitrile molecules. (D) The total force in $z$ direction
and its long-ranged part. The insets show the acetonitrile
charge densities (B) and the total force (D) in
the region of the LV interface. 
Smoothing eliminates the simulation noise still visible
in the bare results. 
The blue lines in (B) and (C) are the same
but on different scales. See Eqs.~\eqref{eqn:conv}, ~\eqref{eqn:aceconv}
and ~\eqref{eq:elecfz}.}
\label{fig:chargeforcesep}
\end{figure*}

\subsection{Electrostatic potential and the long range contribution}
In order to shed light on the importance of the correct treatment of the
long-ranged electrostatics in this interfacial system, we analyze the total equilibrium
charge density $\rho^q_{\rm tot}({\mathbf r})$, comprised of the fixed charges in the silica
substrate and the induced equilibrium mobile charge density
$\rhoq_{\rm ace}(\vect{r})$ in the acetonitrile fluid. The associated electrostatic potentials are given by Poisson's equation (in Gaussian units) as
\begin{equation}
  \V^{\rm tot}(\vect{r}) = \int d\vect{r}^\prime \, \rhoq_{\rm tot}(\vect{r}^\prime )  \cdot
  \frac{1}{\vdiff{r}{r^{\prime}}}.
\label{eqn:smoothedrhomobile}
\end{equation}
and 
\begin{equation}
  \V^{\rm ace}(\vect{r}) = \int d\vect{r}^\prime \, \rhoq_{\rm
ace}(\vect{r}^\prime )  \cdot
  \frac{1}{\vdiff{r}{r^{\prime}}}.
\label{eqn:acesmoothedrhomobile}
\end{equation}

In other work \cite{Chen_Weeks2004,Rodgers_Weeks2008}, we have argued for
the utility of splitting the basic Coulomb $1/r$ interaction
appearing in Eqs.~\eqref{eqn:smoothedrhomobile} and \eqref{eqn:acesmoothedrhomobile} into short- and long-ranged parts based on a ``smoothing length'' $\sigma$ of order a characteristic nearest neighbor spacing:
\begin{equation}
\frac{1}{r} =\vs + \vl=  \frac{{\rm erfc}(r/\sigma)}{r}\  + \frac{{\rm erf}(r/\sigma)}{r}.
\end{equation}
Here \vl\ is the functional form
associated with the electrostatic potential due to a unit Gaussian charge distribution of width
$\sigma$, defined as
\begin{equation}
  \rhoG(\vect{r}) = \frac{1}{\pi^{3/2}\sigma^3} \exp
  \left(-\frac{r^2}{\sigma^2}\right),
  \label{eqn:rhoG}
\end{equation}
implying a \vl\ given by the convolution
\begin{equation}
\vl \equiv \int d\vect{r}^\prime \rhoG(\vect{r^\prime})\cdot \frac{1}{\vdiff{r}{r^{\prime}}}= \frac{\erf(r/\sigma)}{r}~.
  \label{eqn:v1ConvDef}
\end{equation}
As shown in Fig.~\ref{fig:chargeforcesep}(A), \vl\ is slowly-varying in $r$-space over the length scale \sig.

We showed in Ref. \cite{Hu_Weeks2009}
that for bulk acetonitrile the forces from the long-ranged parts of the
Coulomb interactions essentially cancel when \sig\  is properly chosen to be $4${\rm \AA}, and a very accurate description of local pair correlations
arises from considering the simpler ``strong-coupling'' system with only
truncated Coulomb interactions \vs. Moreover, as argued in Ref.\
\cite{Rodgers_Weeks2008,Rodgers_Weeks2008a},
in nonuniform systems, where this force cancellation does
not occur, the charge densities that best reflect the long-ranged
electrostatics are not the bare densities $\rhoq_{\rm tot}(\vect{r})$ and $\rhoq_{\rm ace}(\vect{r})$
appearing in Eqs.~\eqref{eqn:smoothedrhomobile} and \eqref{eqn:acesmoothedrhomobile}
but rather the Gaussian-smoothed charge densities
\begin{equation}
  \rhoqs_{\rm tot}(\vect{r}) \equiv \int d\vect{r}^\prime \,
\rhoq_{\rm tot}(\vect{r}^\prime) \rhoG(\vdiff{r}{r^\prime}),
\label{eqn:conv}
\end{equation}
and 
\begin{equation}
  \rhoqs_{\rm ace}(\vect{r}) \equiv \int d\vect{r}^\prime \,
\rhoq_{\rm ace}(\vect{r}^\prime) \rhoG(\vdiff{r}{r^\prime}),
\label{eqn:aceconv}
\end{equation}
generated from  Eqs.\ (\ref{eqn:smoothedrhomobile}),
(\ref{eqn:acesmoothedrhomobile}) and
(\ref{eqn:v1ConvDef}) when only the long-ranged Coulomb component $v_{1}(\vdiff{r}{r^{\prime}})$ is used:
\begin{eqnarray}
  \V_{1}^{\rm tot}(\vect{r}, \sigma)&\equiv & \int d\vect{r}^\prime \, \rhoq_{\rm tot}(\vect{r}^\prime )  \cdot
  v_{1}(\vdiff{r}{r^{\prime}})  \nonumber \\
& = & \int d\vect{r}^\prime \, \rhoqs_{\rm tot}(\vect{r}^\prime )  \cdot
  \frac{1}{\vdiff{r}{r^{\prime}}}. 
  \label{eqn:smoothedtot}
\end{eqnarray}
and 

\begin{eqnarray}
  \V_{1}^{\rm ace}(\vect{r}, \sigma) &\equiv & \int d\vect{r}^\prime \,
\rhoq_{\rm ace}(\vect{r}^\prime )  \cdot
  v_{1}(\vdiff{r}{r^{\prime}}) \nonumber \\
  & = & \int d\vect{r}^\prime \, \rhoqs_{\rm
ace}(\vect{r}^\prime )  \cdot
  \frac{1}{\vdiff{r}{r^{\prime}}}.
\label{eqn:smoothedace}
\end{eqnarray}
for the total system and the mobile acetonitrile molecules respectively.

Thus we are able to rationalize the
contribution of the long-ranged part of the electrostatics through analysis of
the smooth charge density $\rhoqs_{\rm tot}(\vect{r})$ and the
corresponding electrostatic potential $ \V_{1}^{\rm tot}(\vect{r}, \sigma)$.
One virtue of this perspective is that because of the integration over the smoothing length $\sigma$,
$\rhoqs_{\rm tot}(\vect{r})$ is very slowly varying along the interface in the $x$ and $y$ directions, unlike the bare charge density.
As a result, $\V_{1}^{\rm tot}(\vect{r}, \sigma)=\V_{1}^{\rm tot}(z, \sigma)$
to a very good approximation \cite{Rodgers_Weeks2008,Rodgers_Weeks2008a}
and we can focus only on forces in the $z$-direction given by:
\begin{equation}
 F_z(z, \sigma) =- \frac{\partial \V_{1}^{\rm tot}(z, \sigma)}{\partial z}.
\label{eq:elecfz} \end{equation}

A plot of $\rhoqs_{\rm tot}(z)$, $\rhoqs_{\rm ace}(z)$
and $F_z(z, \sigma)$ is shown in Fig.~\ref{fig:chargeforcesep}.
The magnitude of the acetonitrile charge density close to silica surface
($0<z<10$ {\rm \AA}) is on the order of $0.01$ {\rm e \AA}$^{-3}$ and it
dramatically
reduces to the order of $0.0001$ {\rm e \AA}$^{-3}$ at the LV interface.
The smoothed
charge density in the LS region for the relevant $\sigma=4.0$ {\rm \AA} is much smaller, with
a magnitude of order of only $0.0002$ {\rm e \AA}$^{-3}$.

This implies that the long-ranged ($\sigma=4$ {\rm
\AA}) part of the charge density
only contributes 4\% to the total charge density in the region of LS
interface. Nevertheless, the long-ranged forces play an important role in
properly describing the surface-induced dipole, as discussed in
detail in Ref.\  \onlinecite{Rodgers_Weeks2008} and illustrated below.
However, in the LV interfacial region, the bare charge density is much smaller, and the 
long-ranged ($\sigma=4$ {\rm \AA}) charge density  can contribute as much
as 50\% (see the inset of Fig.~\ref{fig:chargeforcesep} (B)). Analysis of
the force in the $z$ direction (Fig.~\ref{fig:chargeforcesep} (D)) supports
this same conclusion. Therefore, the long-ranged part of the electrostatics
plays a relatively more important role in determining the structure and dynamics of
acetonitrile in the LV interface than near the LS interface, where 
most aspects of the local structure is dominated by short-ranged forces. 

Fig.~\ref{fig:chargeforcesep} (C) shows a comparison of the total smooth
charge density (red) and the smooth charge density of the mobile acetonitrile molecule only
(blue). The overlap of the two densities at distances larger
than $10$ {\rm\AA} indicates complete screening of the induced surface dipole beyond about $10$
{\rm\AA}, consistent with our finding in previous subsections that the dynamics and
structure of acetonitrile in the
region between $25$ {\rm \AA} and $30$ {\rm \AA} have reached the bulk limit.
\section{Conclusion}
\label{sec:con}
We have investigated the structure and dynamics of the acetonitrile system
on silica surface and at its LV interface. The change of melting
point, the slowing down of dynamics for molecules close to silica surface and the tilting
angle of the molecule at the LV interface were all explained in
detail through our studies of structure and dynamics. The antiparallel
correlations in the LS (LV) region was weaker (stronger) than in the bulk.
Both singlet and
collective reorientational time scales in the first layer close to the silica
surface are approximately $6-7$ times slower than the corresponding time
scales in the bulk. The collective reorientational time scale is about 3
times faster than that of the singlet reorientation.  Moreover, we
provide a general explanation of the ambiguity that arises
in determining the orientational profile using interfaces defined with 
different choices of the atoms in a molecular system. 
Our current study of the acetonitrile system provides a general framework for
future MD simulations of LS and LV interfacial systems. We
plan to investigate other typical systems such 
as quadrupolar (carbon dioxide), dipolar (propionitrile) and
associating fluids (hydrogen fluoride and water) on the silica surface and
at their liquid-vapor interfaces.

\section*{Acknowledgment}
The work is supported by a Collaborative Research in Chemistry grant
CHE0628178 from the National Science Foundation. We are grateful
to Prof. John Fourkas and Prof. Robert Walker for the many
stimulating conversations. We also thank Rick Remsing and Jocelyn
Rodgers for comments on the manuscript.


\end{document}